\documentclass[
prx,
superscriptaddress,
twocolumn,
longbibliography
]{revtex4-1}

\usepackage{float}
\usepackage{bbm}
\usepackage{epsfig}
\usepackage{epstopdf}
\usepackage{graphicx, subfigure}
\usepackage{pgfplots}
\usetikzlibrary{patterns}
\usepgfplotslibrary{groupplots}
\usepackage{amssymb}
\usepackage{amsmath}
\usepackage{scalefnt}
\usepackage{color}
\usepackage{hyperref}
\usepackage{qcircuit}
\usepackage[capitalize]{cleveref}
\usepackage{braket,mleftright}
\usepackage{physics}
\usepackage{import}
\usepackage{dcolumn}
\usepackage{csquotes}
\usepackage{tikz}

\floatstyle{plaintop}
\restylefloat{table}

\begin{document}
    \title{Preserving Symmetries for Variational Quantum Eigensolvers in the Presence of Noise}

    \author{George S. Barron}
    \email{gbarron@vt.edu}
    \affiliation{Department of Physics, Virginia Tech, Blacksburg, VA 24061, U.S.A}
    \author{Bryan T. Gard}
    \affiliation{Department of Physics, Virginia Tech, Blacksburg, VA 24061, U.S.A}
    \author{Orien J. Altman}
    \affiliation{Department of Physics, Virginia Tech, Blacksburg, VA 24061, U.S.A}
    \author{Nicholas J. Mayhall}
    \affiliation{Department of Chemistry, Virginia Tech, Blacksburg, VA 24061, U.S.A}
    \author{Edwin Barnes}
    \affiliation{Department of Physics, Virginia Tech, Blacksburg, VA 24061, U.S.A}
    \author{Sophia E. Economou}
    \affiliation{Department of Physics, Virginia Tech, Blacksburg, VA 24061, U.S.A}

    \date{\today}

    \begin{abstract}
        One of the most promising applications of noisy intermediate-scale quantum computers is the simulation of molecular Hamiltonians using the variational quantum eigensolver. We show that encoding symmetries of the simulated Hamiltonian in the VQE ansatz reduces both classical and quantum resources compared to other, widely available ansatze. Through simulations of the H$_2$ molecule, we verify that these improvements persist in the presence of noise. This simulation is performed with IBM software using noise models from real devices. We also demonstrate how these techniques can be used to find molecular excited states of various symmetries using a noisy processor. We use error mitigation techniques to further improve the quality of our results.
    \end{abstract}

    \maketitle
    
    \section{Introduction}\label{sec:introduction}

        Quantum computers are believed to be one of the most promising technologies currently being developed that will help extend the reach of scientific discovery. This may be achieved through quantum simulation \cite{Feynman1982}, which leverages the properties of a quantum processing unit (QPU) to simulate naturally occurring quantum mechanical systems. One of the most popular algorithms within the reach of near term devices is the variational quantum eigensolver (VQE) \cite{Peruzzo2014, OMalley2016, McClean2016, kandala2017hardware, Colless2018, Preskill2018}.
        
        This algorithm falls under a more general class of algorithms known as hybrid variational quantum algorithms \cite{farhi2014quantum, pagano2019quantum, romero2019variational, leyton2019robust, Zhueaaw9918, benedetti2019generative, verdon2019learning}. The general principle of these algorithms is to use a feedback loop between the quantum and classical computers to minimize a predefined cost function. This method has been applied to a variety of quantum systems in both theoretical \cite{Kivlichan2018, PhysRevA.99.062304, PhysRevApplied.11.044087, PhysRevLett.122.230401, anschuetz2019variational, Moll_2018, Barkoutsos2018, gard2019efficient, doi:10.1002/qute.201900070, barkoutsos2019improving, robert2019resource, PhysRevA.99.012334}, and experimental \cite{OMalley2016, kandala2017hardware, larose2019variational, Zhueaaw9918, PhysRevLett.120.210501, PhysRevA.98.032331, Peruzzo2014, kokail2019self} contexts. In the case of the VQE, the predefined function is the expectation value of the simulated Hamiltonian with respect to the state of the QPU. Additionally, a variety of techniques are available that use the VQE to find higher excited states of such systems \cite{Colless2018, Higgott2019variationalquantum, ollitrault2019quantum}.
        
        The advantage of hybrid variational algorithms lies in their ability to exploit advanced classical computational resources without having to store the wavefunction on a classical computer. Wavefunctions are instead prepared and measured on the QPU. The success of these algorithms depends sensitively on the complexity of the quantum circuits used to prepare the wavefunctions.
        This in turn is fundamentally related to the number of parameters needed to describe arbitrary states. Hence, a fundamental question is ``how likely is it that a given parameterized quantum circuit (a variational ansatz) will represent the targeted state with sufficient accuracy?'' Typically there are no accuracy guarantees for a given ansatz.
        One approach to designing ansatze with tunable accuracy, is to iteratively build a customized variational ansatz in a ``problem tailored'' fashion, as proposed with the ADAPT-VQE algorithm \cite{Grimsley2018, tang2019qubitadaptvqe}. This is accomplished by growing the ansatz operator-by-operator, each time selecting the operator which creates the largest change to the objective function, using the VQE as a subroutine at each step.
        This is in contrast to alternative approaches that fix the number of parameters from the start. In Ref.~\cite{gard2019efficient}, we showed how to create quantum circuits that have exactly the number of variational parameters necessary to describe any state in the relevant symmetry subspace of the Hilbert space.
        
        One of the advantages of Ref.~\cite{gard2019efficient} is that the ansatz is guaranteed to find the correct state in the absence of noise, both for ground and for excited states.
        Other algorithms that accomplish excited state simulation are sometimes iterative and rely on previous results of a VQE, thus introducing computational overheads.
        Despite these advantages of symmetry-enforcing circuits, however, it is not clear how the presence of noise might affect their performance. 
        The operators which create noise clearly will not commute with the same symmetries as the target Hamiltonian, 
        and as such, might drive the system into undesirable symmetry subspaces. 
        Will a circuit that preserves these symmetries perform worse due to a lack of relevant error mitigating degrees of freedom?
        More generally,
        among the outstanding challenges of VQEs are issues that involve optimizing a rapidly varying function with many parameters, limited computational resources, and in the presence of noise \cite{Kuhn2018, McClean2018, Grant2019initialization, parrish2019jacobi, parrish2019hybrid, verteletskyi2019measurement, barkoutsos2019improving, doi:10.1021/acs.jctc.9b00791, gokhale2019minimizing, wilson2019optimizing, bonet2019nearly, izmaylov2019revising}. Work has been done to manipulate the input and output of the QPU so as to mitigate and characterize the error \cite{Bonet2018, McArdle2018, PhysRevA.100.010302, PhysRevX.10.011004}. 
        
        In this work, we consider the effect that noise has on the ability of a VQE to preserve symmetries. The simulations here are done using IBM's Qiskit \cite{qiskit} software, using noise models derived from real superconducting devices. We perform these simulations using our previously developed techniques \cite{gard2019efficient} for encoding symmetries of the molecular Hamiltonians directly into the state preparation circuit. We find that our ans\"atze can outperform standard, ad hoc ans\"atze. We adapt this approach to solving for excited states that obey certain symmetries. This allows us to guarantee the ability of the ansatz to find the desired state, and in fact, accomplish this with a minimal number of variational parameters. Additionally, our approach for producing excited states does not rely on previous runs of VQEs and is parallelizable for the different excited states calculated.
        
        This paper is structured as follows.
        In Sec.~\ref{sec:background} we provide the necessary background to make this paper self contained.
        Specifically, in Sec.~\ref{sec:vqe}, we review the theory behind VQE and its application to quantum chemistry problems.
        In Sec.~\ref{sec:symans} we review our earlier work on symmetry preserving VQE ans\"atze.
        In Secs.~\ref{sec:noiseless} and \ref{sec:effects_of_noise}, we perform simulations on the $\text{H}_2$ molecule and analyze the performance of these algorithms in noiseless and noisy simulation contexts, respectively. In Sec.~\ref{sec:effects_of_noise} we use, benchmark, and compare a variety of error mitigation techniques, as well as discuss challenges associated with noisy optimization.
        In Sec.~\ref{sec:symm_in_noisy} we compare the ability of the symmetry preserving and ad hoc ans\"atze to preserve desired symmetries in a noisy environment.
        In Sec.~\ref{sec:excited} we present the results of this procedure for excited state simulation.
        In Sec.~\ref{sec:conc}, we give a summary of our conclusions.

    \section{Background}\label{sec:background}
    
        \subsection{Variational Quantum Eigensolver}\label{sec:vqe}
        The goal of a VQE is to minimize an objective function of the form $f(\vec{\theta}) = \bra{\psi(\vec\theta)}H\ket{\psi(\vec\theta)}$, where $\ket{\psi(\vec\theta)}= U(\vec{\theta})\ket{\psi_0}$ is the state of the QPU after initializing the register to some initial state $\ket{\psi_0}$ and applying a circuit with unitary $U(\vec{\theta})$ where $\vec{\theta}$ is a vector of variational parameters chosen by an optimization algorithm. Per the variational principle,  $f(\vec{\theta}) \geq E_0$ where $E_0$ is the ground state energy of the Hamiltonian $H$.
        
        We focus on the second quantized molecular Hamiltonian \begin{equation}\label{chemham}\hat{H}_f=\sum_{ij}h_{ij}\hat{a}^\dagger_i\hat{a}_j+\sum_{ijkl}g_{ijkl}\hat{a}^\dagger_i\hat{a}^\dagger_j\hat{a}_k\hat{a}_l,\end{equation} where $\hat{a}$ ($\hat{a}^\dagger$) is the fermionic annihilation (creation) operator. The quantities $h_{ij}$  and $g_{ijkl}$ are the single and double electron integrals respectively, which can be computed efficiently on a classical computer with existing software. To evaluate the objective function $f(\vec{\theta})$ on the QPU, one must map the Fermionic operators to a set of qubit operators in a way that accounts for the Fermionic anti-commutation relations. This can be accomplished with several different methods, including the Jordan-Wigner (JW) \cite{Jordan1928}, parity \cite{doi:10.1063/1.4768229}, Bravyi-Kitaev \cite{BRAVYI2002210}, or BKSF \cite{PhysRevResearch.1.033033} mappings. In this work we choose the JW mapping as it directly maps occupations of fermionic spin orbitals to excitations of qubits. This choice makes it far simpler to impose symmetries such as particle number and spin projection, due to the fact that these operators are local in the JW mapping. The resulting Hamiltonian appearing in the objective function $f(\vec{\theta})$ is a weighted sum of Pauli strings $\hat{H}=\sum_{i}\alpha_{i} T_i$ where $\alpha_i \in \mathbb{R}$ are the weights of Pauli strings and $T_i \in \{I,X,Y,Z\}^{\otimes n}$ for $n$ qubits (with the JW mapping, this is the number of spin orbitals). The role of the QPU is to compute expectation values of this Hamiltonian. In particular, for a given set of parameters $\vec{\theta}$, the QPU calculates $\left\{ \left< T_i \right>_{\vec{\theta}} \right\}_i$ which may then be weighted on the classical computer according to the coefficients $\alpha_i$ to form $\left< H \right>_{\vec{\theta}}$. Work has been done to reduce the number of measurements at this step based on the properties of $\{ T_i \}_i$ \cite{doi:10.1021/acs.jctc.9b00791}.
        
        Despite substantial progress, accurately performing VQEs on current and near-term hardware is still challenging for a number of reasons. Firstly, calls to the objective function $f(\vec{\theta})$ are inherently probabilistic, meaning one must use an optimization algorithm to minimize $f(\vec{\theta})$ that is resistant to noise. Additionally, the number of variational parameters used to construct the variational form can potentially limit the performance of an optimization algorithm in finding a suitable minimum for the estimated ground state energy. Furthermore, the depth of circuits and number of CNOT gates are costly resources when executing circuits on a real QPU, so creating ans\"atze that efficiently use these resources is paramount.

    \subsection{Symmetry Preserving Ans\"atze}\label{sec:symans}
    
        The minimal number of parameters for a general Hilbert space of $n$ qubits (i.e. $n$ spin orbitals) is given by $2(2^n-1)$. However if symmetries are present, they can be used to reduce this parameter count, sometimes dramatically. For example, if particle number is conserved, then the number of parameters reduces to $2(\binom{n}{m}-1)$, where $m<n$ is the number of fermions. If the Hamiltonian also respects time-reversal symmetry, as is often the case in molecular problems, then the energy eigenstates can be chosen to have only real coefficients, which further reduces the parameter count by a factor of 2. Total spin $s$ and spin projection $s_z$ are also commonly conserved, in which case the parameter count becomes \cite{gard2019efficient} \begin{equation}\label{dimspin}\begin{split}&\sum_{k=0}^{m/2-s}\binom{n/2}{k}\binom{n/2-k}{m-2k}\\&\times\frac{(2s+1)(m-2k)!}{(m/2-k-s)!(m/2-k+s+1)!}.\end{split}\end{equation}
            Choosing a particular symmetry-preserving ansatz amounts to restricting the search space of the optimizer to a smaller region of the Hilbert space. This in turn reduces the number of variational parameters, lessening the burden on the classical optimizer. We show below that it also lessens the demands on the QPU by cutting down the number of function calls in the optimization process and reducing the circuit depth.
            
        Using a symmetry-preserving ansatz contrasts sharply with using ans\"atze that rely on ad hoc arguments \cite{kandala2017hardware, PhysRevApplied.11.044092}. These ans\"atze usually attempt to express all possible states in the Hilbert space, including the ground state of the Hamiltonian in question. The conventional wisdom here is that more variational parameters result in more flexibility in producing the target states. However, too many variational parameters can result in large circuit depths and challenging conditions for the optimizer.  It may also not be possible to exactly represent the ground state with these types of ans\"atze if they are not designed to uniformly cover the relevant part of the Hilbert space.
            
\begin{figure}[!tb]
            \[ \Qcircuit @C=1.25em @R=1em {
            \ket{0} &	&	\qw	&	\multigate{1}{A}	&	\qw	&	\qw	&		&		&		&	\qw	&	\multigate{1}{A}	&	\qw	&	\qw	\\
            \ket{0} &	&	\gate{X}	&	\ghost{A}	&	\multigate{1}{A}	&	\qw	&		&		&		&	\qw	&	\ghost{A}	&	\multigate{1}{A}	&	\qw	\\
            \ket{0} &	&	\qw	&	\qw	&	\ghost{A}	&	\qw	&		&		&		&	\qw	&	\qw	&	\ghost{A}	&	\qw	\\
	        &		&		&		&		&		&		&		&		&		&		&		\\
            \vdots	&		&		&	\vdots	&		&		&	\ddots	&		&		&		&		&	\vdots	\\
	        &		&		&		&		&		&		&		&		&		&		&		\\
            \ket{0} &	&	\gate{X}	&	\qw	&	\multigate{1}{A}	&	\qw	&		&		&		&	\qw	&	\qw	&	\multigate{1}{A}	&	\qw	\\
            \ket{0} &	&	\qw	&	\multigate{1}{A}	&	\ghost{A}	&	\qw	&		&		&		&	\qw	&	\multigate{1}{A}	&	\ghost{A}	&	\qw	\\
            \ket{0} &	&	\gate{X}	&	\ghost{A}	&	\qw	&	\qw	&		&		&		&	\qw	&	\ghost{A}	&	\qw	&	\qw
            } \]
            \caption{Circuit used to produce the ``ASWAP" ansatz. Each $A$ gate has two separate parameters, $\theta$ and $\phi$. Each $A$ gate preserves the particle number of the state it acts on, so the entirety of the ansatz does too. The initial $X$ gates are used to place the register in the desired particle number subspace.}
            \label{fig:circ}
            \end{figure}
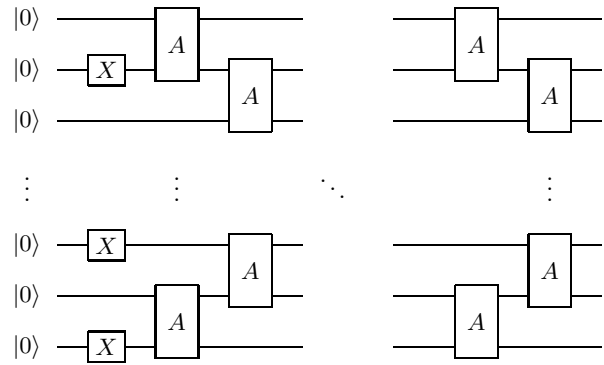

            In this work, we primarily focus on the ``ASWAP'' ansatz, which is constructed using gates that preserve the number of excitations in a state. In the case of a molecular Hamiltonian, this corresponds to fixing the number of electrons occupying some number of orbitals. The gates we consider \cite{Barkoutsos2018} are of the form 
            \begin{eqnarray}
                A(\theta ,\phi )=\begin{pmatrix}
                1 & 0 & 0 & 0\\
                0 & \cos \theta  & e^{i\phi }\sin \theta  & 0\\
                0 & e^{-i\phi }\sin \theta & -\cos \theta & 0\\
                0 & 0 & 0 &1
            \end{pmatrix}.\label{eq:Agate}
            \end{eqnarray}
            Ref.~\cite{gard2019efficient} showed how to construct circuits that preserve particle number and spin projection $S_z$ by tiling these $A$ gates in a regular pattern like the one shown in Fig.~\ref{fig:circ}. Preserving any value of $\left< S_z \right>$ is accomplished by dividing the register of qubits into two halves (the upper half represents spin-up orbitals, and the lower half spin-down). The parameters for the $A$ gates that join the two spin subspaces are then set to $0$. The desired $S_z$ subspace can be selected by changing which subset of qubits are acted on by the initial $X$ gates. Further details of this procedure can be found in Ref.~\cite{gard2019efficient}. In this work, we primarily use the ASWAP ansatz that fixes $\left< S_z \right> = 0$, except when considering excited states in Sec.~\ref{sec:excited}.
            
            \section{Performance comparisons of various ans\"atze without noise}\label{sec:noiseless}
            
            Before using the symmetry preserving ans\"atze in noisy contexts, we first demonstrate that the correct ground state can be prepared in a noiseless context. To do this, we run different instances of the VQE algorithm for $\text{H}_2$ under the JW mapping on four qubits using a noiseless simulator. The noiseless simulator used is the ``statevector simulator'' implemented in Qiskit. To perform the optimization of the variational parameters in the VQE, we use the limited memory BFGS optimization algorithm. For noiseless simulations, the number of CNOT gates in an ansatz will not affect the quality of the results. Thus, if a given ansatz is able to reach chemical accuracy, the relevant benchmark is how many function evaluations are required to do so.
            
            Because the specific form of our ASWAP ans\"atze depends on the symmetry subspace we are targeting, we must specify this subspace upfront. In the case of $\text{H}_2$, we expect a singlet ground state, so we have $\left< N \right> = 2$ and $\left< S_z \right> = 0$.
            The ans\"atze that we compare against are standard ad hoc ans\"atze available in Qiskit \cite{qiskit}. Generally, the structure of these ans\"atze involves interleaving layers of single-qubit rotations with entangling operations. In the case of the RY(RYRZ) ans\"atze, layers of single-qubit $R_Y(\theta)$ ($R_Y(\theta) R_Z(\phi)$) operations on all qubits are interleaved with CZ gates. However, since CZ and CNOT are locally equivalent, we instead count the number of CNOT operations for comparison. The number of such layers in the ansatz is referred to as the depth. The SwapRZ ansatz interleaves $R_Z(\theta)$ operations with parameterized SWAP operations, meaning that this ansatz preserves particle number symmetry. In contrast with our ASWAP ansatz, the SwapRZ ansatz does not in general span the full symmetry subspace, and so does not necessarily capture the ground state exactly. Additionally, the ASWAP ansatz generally has fewer parameters and CNOTs. In these ad hoc ans\"atze, the entangling operations performed at each layer can be chosen for the appropriate problem. For the ans\"atze considered here, we perform entangling operations between all pairs of qubits unless specified otherwise.
            
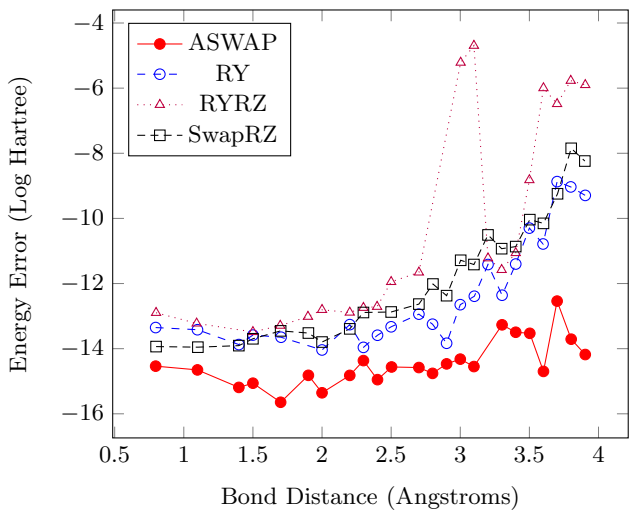
\begin{figure}
                \centering
                \begin{tikzpicture}
                    \begin{axis}[
                        ylabel={Energy Error (Log Hartree)},
                        xlabel={Bond Distance (Angstroms)},
                	    legend pos=north west
                        ]
                        
                        \addplot[red, solid, mark=*, mark options={solid}] table [x=x, y=y, col sep=comma] {noiseless_aswap.csv};
                        \addplot[blue, dashed, mark=o, mark options={solid}] table [x=x, y=y, col sep=comma] {noiseless_ry.csv};
                        \addplot[purple, dotted, mark=triangle, mark options={solid}] table [x=x, y=y, col sep=comma] {noiseless_ryrz.csv};
                        \addplot[black, densely dashed, mark=square, mark options={solid}] table [x=x, y=y, col sep=comma] {noiseless_swaprz.csv};
                        
                	    \legend{ASWAP,RY,RYRZ, SwapRZ}
                        
                    \end{axis}
                \end{tikzpicture}
                \caption{Results for $\text{H}_2$ in a noiseless simulator with the limited memory BFGS algorithm. All results are well below chemical accuracy and sufficiently converged. To account for bad initial variational parameters, each result was obtained by running additional optimizations with randomized initial variational parameters. The lowest energy found with this method was then chosen as the final result. The median number of iterations of the BFGS optimizer for each ansatz were: ASWAP 414, RY 2142, SwapRZ 3955, RYRZ 4851.}
                \label{fig:noiseless}
            \end{figure}

            In Fig.~\ref{fig:noiseless}, we illustrate the performance of several ad hoc ans\"atze, along with our ASWAP ansatz. All the ans\"atze considered are able to find the correct ground state energy to well below chemical accuracy ($\sim1.5$~milliHartree). One distinction between our ansatz and the ad hoc ans\"atze, however, is that the ASWAP ansatz requires at least five times fewer function evaluations to converge compared to any of the other considered ans\"atze, while simultaneously having lower error rates. For the example shown in Fig.~\ref{fig:noiseless}, ASWAP requires 414 function calls, while RY uses 2142. This improvement is expected since the number of variational parameters is smaller than the other considered ans\"atze. This illustrates the key point that enforcing known symmetries of the simulated Hamiltonian can significantly reduce the computational load on the QPU, namely the number of calls to the objective function.

\begin{figure}
                \centering
                \begin{tikzpicture}
	             \begin{axis}[%
	             scatter/classes={%
		              ASWAP={mark=*,red},%
		              RY={mark=o,blue},%
                		RYRZ={mark=triangle,draw=purple},
                		SwapRZ={mark=square,draw=black}
                	},
                	minor tick num=5,
                	legend pos=north west,
                	xlabel={\# Parameters},
                	ylabel={\# CNOTs}
                	]
                	\addplot[scatter,only marks, scatter src=explicit symbolic]%
	             table[meta=label] {
                    x	y	label
                    1	0	ASWAP
                    1	0	ASWAP
                    1	0	ASWAP
                    1	1	ASWAP
                    3	6	ASWAP
                    1	1	ASWAP
                    8	3	RY
                    16	9	RY
                    12	6	RY
                    32	9	RYRZ
                    24	6	RYRZ
                    16	3	RYRZ
                    14	24	SwapRZ
                    34	72	SwapRZ
                    24	48	SwapRZ
                	};
                	\legend{ASWAP,RY,RYRZ, SwapRZ}
                	\end{axis}
                \end{tikzpicture}
                \caption{Comparison of resources used in different types of variational forms. Ans\"atze with more parameters require more optimization steps and hence calls to the objective function. Ans\"atze with more CNOT gates suffer from higher error rates and are more limited by decoherence. The data displayed include instances of RY(RZ) ans\"atze where the entanglement layer consists of CZ gates between nearest neighbors. The circuit depth of each family is around ~10-35 operations depending on the characteristics of the ansatz, with the exception of the SwapRZ ans\"atze which tend to have longer depths (from ~50-200 operations).}
                \label{fig:resources}
            \end{figure}
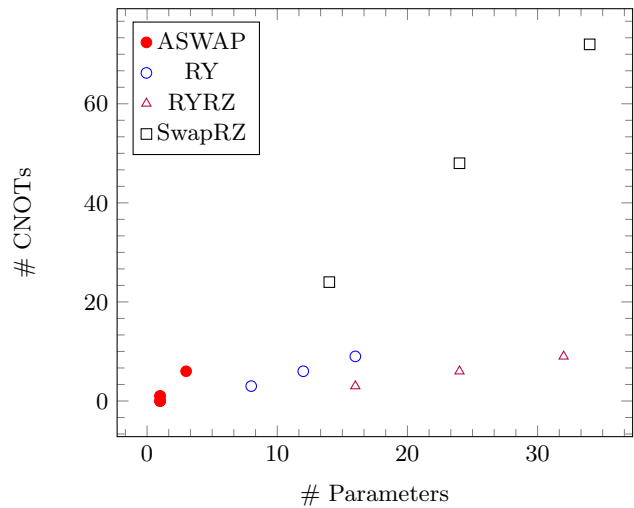

            The metrics that we use to evaluate the resource ``efficiency" of a particular ansatz are the number of variational parameters and the number of CNOT operations it contains. We summarize these resource requirements for ASWAP and for several of the ad hoc ans\"atze described above in Fig.~\ref{fig:resources}. Generally, as expected, the number of variational parameters in the ASWAP ans\"atze can be small, despite being able to quickly and accurately prepare the correct ground state.

             \section{Effects of noise on performance of VQE with various ans\"atze}\label{sec:effects_of_noise}

            \subsection{Hardware Noise and Noise Model Simulation}
            
            All of the ans\"atze used in this paper are applicable to any general-purpose quantum processor. Nevertheless, for the purposes of noisy simulation, we will consider a noise model that is derived from information about the IBMQ processors. In particular, we include information from the devices about measurement errors, single- and two-qubit gate errors, depolarization, and thermal relaxation errors that are dependent on individual gate times.
	        
	        To establish a vocabulary, the different simulators in Qiskit that we use here are as follows. We refer to the noiseless simulator that keeps track of the state vector of the register under unitary operations as the ``state vector simulator''. We refer to the noisy simulator that includes the above information as the ``QASM simulator''. When using the QASM simulator, we will use noise models constructed from information about the Vigo, Boeblingen, Ourense, and Johannesburg IBM Q devices.

            \subsection{Error Mitigation Techniques} \label{sec:error_mit}
            
            When running a VQE in the presence of noise, we use three error mitigation techniques. The first of these addresses state preparation and measurement (SPAM) errors by computing and inverting a matrix $T$ of dimension $2^n \times 2^n$ whose entries represent the probability of preparing one state and immediately measuring another. Inverting this matrix allows one to perform error mitigation on population counts of a particular experiment, thereby accounting for SPAM errors \cite{PhysRevLett.114.200501, sun2018efficient}. Here, we use the implementation in Qiskit.
            
             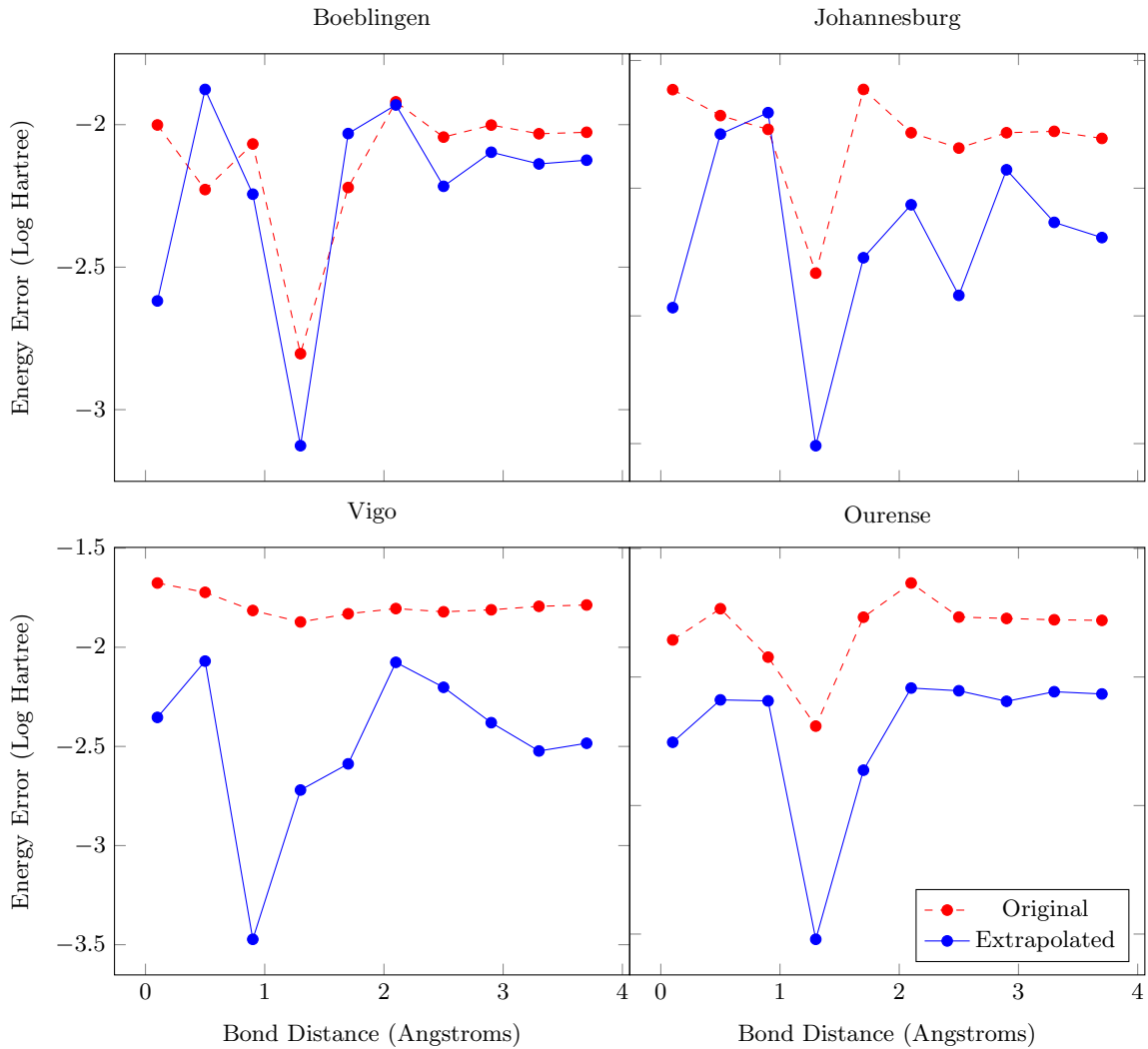
\begin{figure*}
                \centering
                \begin{tikzpicture}
                    \begin{groupplot}[
                        group style = {group size = 2 by 2,
                        x descriptions at=edge bottom,
                        y descriptions at=edge left,
                        horizontal sep = 0pt,
                        vertical sep = 25pt
                        },
                        xlabel={Bond Distance (Angstroms)},
                        ylabel={Energy Error (Log Hartree)}
                        ]
                    
                    \nextgroupplot[title=Boeblingen]
                    \addplot[red, dashed, mark=*, mark options={solid}] table [x=distance, y=abs_log_err, col sep=comma] {extrap_data_boeblingen_False.csv};
                    \addplot[blue, solid, mark=*, mark options={solid}] table [x=distance, y=abs_log_err, col sep=comma] {extrap_data_boeblingen_True.csv};
                    
                    \nextgroupplot[title=Johannesburg]
                    \addplot[red, dashed, mark=*, mark options={solid}] table [x=distance, y=abs_log_err, col sep=comma] {extrap_data_johannesburg_False.csv};
                    \addplot[blue, solid, mark=*, mark options={solid}] table [x=distance, y=abs_log_err, col sep=comma] {extrap_data_johannesburg_True.csv};
                    
                    \nextgroupplot[title=Vigo]
                    \addplot[red, dashed, mark=*, mark options={solid}] table [x=distance, y=abs_log_err, col sep=comma] {extrap_data_vigo_False.csv};
                    \addplot[blue, solid, mark=*, mark options={solid}] table [x=distance, y=abs_log_err, col sep=comma] {extrap_data_vigo_True.csv};
                    
                    \nextgroupplot[legend pos=south east, title=Ourense]
                    \addplot[red, dashed, mark=*, mark options={solid}] table [x=distance, y=abs_log_err, col sep=comma] {extrap_data_ourense_False.csv};
                    \addplot[blue, solid, mark=*, mark options={solid}] table [x=distance, y=abs_log_err, col sep=comma] {extrap_data_ourense_True.csv};
                    
                    \legend{Original, Extrapolated}
                    \end{groupplot}
                \end{tikzpicture}
                \caption{Comparison of results with and without Richardson extrapolation for $\text{H}_2$ at varying interatomic distances for the ASWAP ansatz. Some points on the extrapolated curve are within chemical accuracy ($\sim1.5$~milliHartree). Richardson extrapolation provides as much two orders of magnitude improvement over a standard VQE run.}
                \label{fig:richardson_extrapolation}
            \end{figure*}
            
            The second technique, known as Richardson extrapolation \cite{PhysRevLett.119.180509, PhysRevLett.120.210501, kandala2019error}, involves systematically increasing the amount of error in a given computation in a controlled way so that one can then extrapolate an objective function to a point at which those errors would be reduced. In this case, the extrapolation occurs by taking each CNOT gate in a particular circuit and inserting an additional even number of CNOTS immediately after it. For instance, each CNOT gate in a given circuit would then become a sequence of an odd number of CNOT gates. In this way the logic of the circuit is unaffected, but in the presence of noise and faulty gates, more redundant CNOT gates will increase the error in the results. In Fig.~\ref{fig:richardson_extrapolation}, this procedure is shown for a variety of bond distances. In this case, we find that the extrapolated minimum energy obtained by the VQE is more accurate than the original results by as much as an order of magnitude in the energy error. Moreover, the extrapolated results are within a standard deviation of chemical accuracy for most of the interatomic distances.
            
            The third error mitigation we use is partial symmetry enforcement during measurement. This noise mitigation relies on the fact that the true ground state has a known, definite particle number and therefore the measurement results should also have this symmetry, even in the presence of noise. This mitigation is only partially possible since the measurement process itself requires post-rotation gates, which do not themselves conserve particle number. However, for the chemical Hamiltonians we are interested in, we can always group together Pauli strings which do preserve particle number. For instance, $\text{H}_2$ under the JW mapping onto 4 qubits can be divided into 5 sets of commuting Pauli strings; one of these sets is made up of only Pauli strings that commute with the total particle number operator (terms like $IIIZ,ZZII,IZII$ etc.). During the measurement of this set, we can enforce that the particle number must match that of the ground state, discarding results which do not respect this symmetry. This process simply post-selects data which we know to positively contribute to our goal of finding the ground state. The measurements of Pauli strings that do not commute with the total particle number operator are not modified.
            
\begin{figure}
                \centering
                \begin{tikzpicture}
                    \begin{axis}[
                        ybar,
                        ylabel={Energy Error (Hartree)},
                        symbolic x coords = {None, RE, Sy, Spam, SpamRE, SpamSy, SpamSyRE},
                        x tick label style={rotate=90,anchor=east},
                        yticklabel style={/pgf/number format/fixed,/pgf/number format/precision=5},
                        scaled y ticks=false,
                        ytick={0.0, 0.02, 0.04, 0.06, 0.08, 0.10, 0.12, 0.14}
                        ]
                        \addplot[blue, fill=blue, opacity=0.4] coordinates {(None, 0.130542421) (RE, 0.090362743) (Sy, 0.015195239) (Spam, 0.070273964) (SpamRE, 0.013527336) (SpamSy, 0.004873408) (SpamSyRE, -0.001257111)}; 
                        \addplot[red, fill=red, opacity=0.4, postaction={pattern=north east lines}] coordinates {(None, 0.086351051) (RE, 0.084808758) (Sy, 0.019359942) (Spam, 0.013392721) (SpamRE, -0.004547497) (SpamSy, 0.011784138) (SpamSyRE, 0.00614709)}; 
                        \legend{RY, ASWAP}
                    \end{axis}
                \end{tikzpicture}
                \caption{Performance of two competing ans\"atze with different combinations of error mitigation strategies for $\text{H}_2$ at equilibrium. Each of these strategies is introduced in Sec.~\ref{sec:error_mit}. The abbreviations for the strategies are RE (Richardson Extrapolation), Sy (Partial Symmetry Enforcement), Spam (SPAM), or combinations of any of these. }
                \label{fig:comb_strats}
            \end{figure}
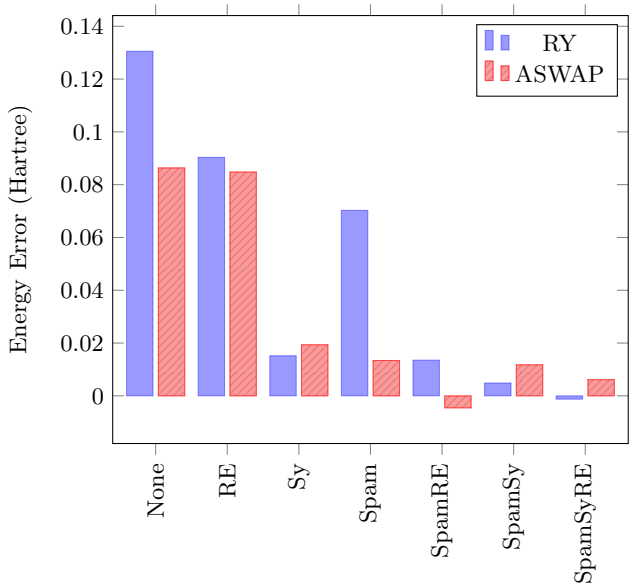

            We employ all of these strategies in various combinations in a noisy H$_2$ simulation using both the RY and ASWAP ans\"atze. The results are shown in Fig.~\ref{fig:comb_strats}. We find that the largest improvement in performance for the RY and ASWAP ans\"atze come from partial symmetry enforcement and SPAM strategies, respectively. In the case of RY, the large improvement under partial symmetry enforcement is likely due to the fact that RY can prepare states that do not obey the desired symmetry, but partial symmetry enforcement corrects this by classical post-processing. In the case of ASWAP, the large improvement from SPAM is likely due to the structure of the ansatz. Specifically, the ASWAP ansatz uses logic that assumes a fixed initial state of the register. Moreover, measurement errors can in principle violate the expected value of the particle number. Both of these effects can be suppressed by SPAM error mitigation.

            \subsection{Noisy Optimization}
            
            Minimization of the objective function $f(\vec\theta)$ requires an optimization algorithm that is resistant to noise. In previous works, variational quantum algorithms have been implemented using the SPSA \cite{spall1992multivariate}, ADAM \cite{kingma2014adam}, COBYLA \cite{Powell1994}, and DIRECT \cite{Jones1993} algorithms. For this work, we choose the DIRECT  and COBYLA algorithms for three primary reasons. First, the DIRECT algorithm has been demonstrated to be resistant to substantial amounts of noise in variational quantum contexts \cite{kokail2019self}. Second, DIRECT is a global optimization algorithm that rarely gets trapped in local minima. In this sense, DIRECT sidesteps the need to consider carefully chosen (and potentially repeated) initial conditions for the variational parameters. Finally, though DIRECT does not converge quickly, it gets within the vicinity of the correct solution quickly. This trade-off is beneficial for the case of a noisy objective function since high levels of precision in the objective function are limited by statistical and hardware errors. We also choose the COBYLA optimization algorithm for similar reasons, namely its ability to find an accurate estimate of the ground state in a short number of function calls and resilience to noise. In previous experiments, it has been shown that the number of shots limits the resulting accuracy of the final energy estimate~\cite{kandala2017hardware}. For this reason, we performed the maximum number of shots allowed by IBM devices (8192). 
            
            For our noisy simulations, the backend chosen is the QASM simulator, which uses a noise model constructed to mimic a variety of IBMQ quantum processors. We use the DIRECT and COBYLA optimization algorithms. In all simulations using DIRECT, we limit the optimization to a budget of $100$ or $300$ calls to the objective function. For the COBYLA optimizer, on average the optimizer converges with about $400$ calls to the objective function.
            
            \subsection{Results for noisy VQE simulations}
            
            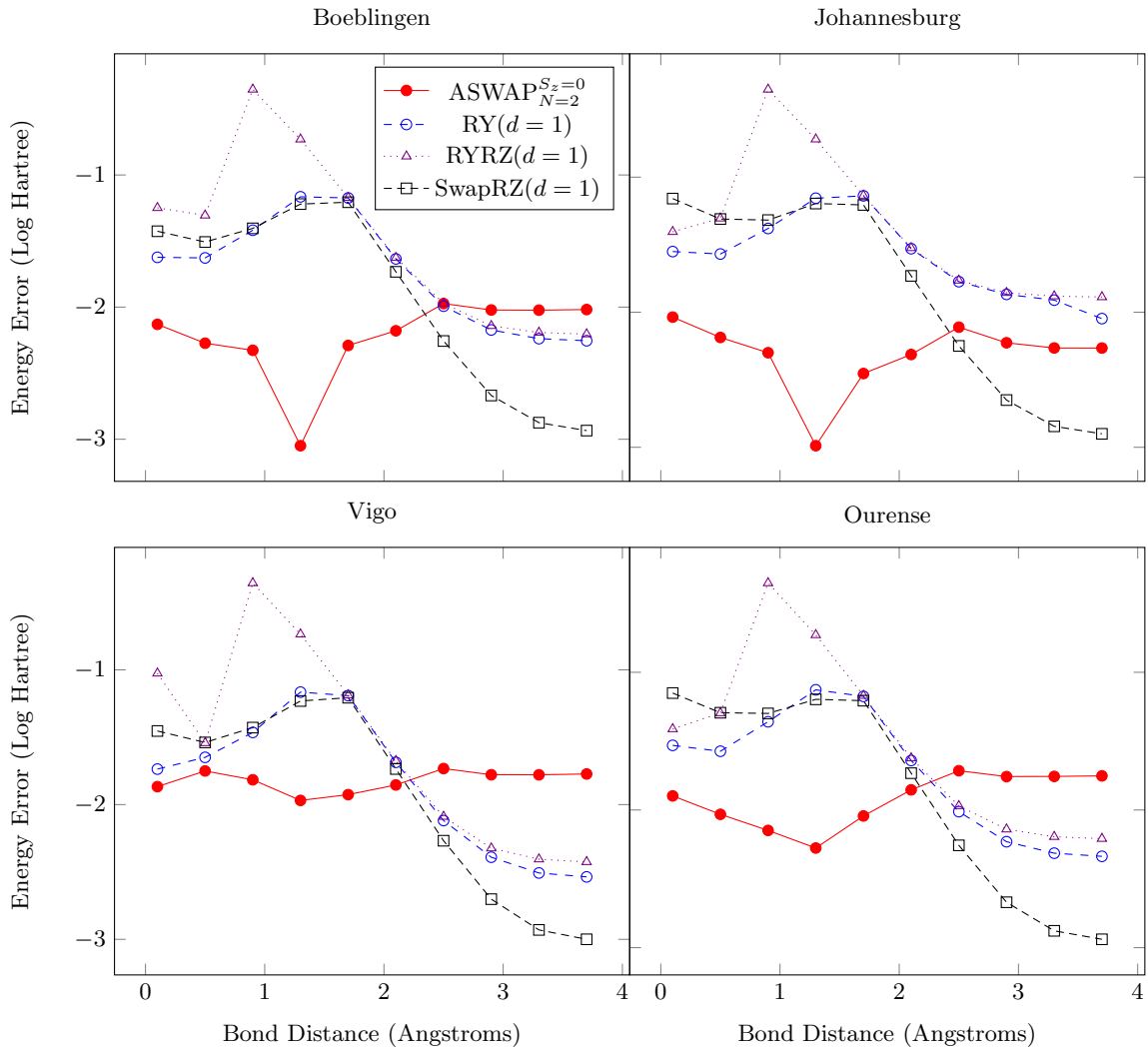
\begin{figure*}
                \centering
                \begin{tikzpicture}
                    \begin{groupplot}[
                        group style = {group size = 2 by 2,
                        x descriptions at=edge bottom,
                        y descriptions at=edge left,
                        horizontal sep = 0pt,
                        vertical sep = 25pt
                        },
                        xlabel={Bond Distance (Angstroms)},
                        ylabel={Energy Error (Log Hartree)},
                        legend pos=north east
                        ]
                    
                    \nextgroupplot[title=Boeblingen]
                    \addplot[red, solid, mark=*, mark options={solid}] table [x=distance, y=abs_log_err, col sep=comma] {comparison_boeblingen_ASWAPn40m2S_z00.csv};
                    \addlegendentry{$\text{ASWAP}_{N=2}^{S_z=0}$}
                    \addplot[blue, dashed, mark=o, mark options={solid}] table [x=distance, y=abs_log_err, col sep=comma] {comparison_boeblingen_RYd1connfull.csv};
                    \addlegendentry{$\text{RY}(d=1)$}
                    \addplot[violet, dotted, mark=triangle, mark options={solid}] table [x=distance, y=abs_log_err, col sep=comma] {comparison_boeblingen_RYRZd1connfull.csv};
                    \addlegendentry{$\text{RYRZ}(d=1)$}
                    \addplot[black, densely dashed, mark=square, mark options={solid}] table [x=distance, y=abs_log_err, col sep=comma] {comparison_boeblingen_SwapRZd1connfull.csv};
                    \addlegendentry{$\text{SwapRZ}(d=1)$}
                    
                    \nextgroupplot[title=Johannesburg]
                    \addplot[red, solid, mark=*, mark options={solid}] table [x=distance, y=abs_log_err, col sep=comma] {comparison_johannesburg_ASWAPn40m2S_z00.csv};
                    \addplot[blue, dashed, mark=o, mark options={solid}] table [x=distance, y=abs_log_err, col sep=comma] {comparison_johannesburg_RYd1connfull.csv};
                    \addplot[violet, dotted, mark=triangle, mark options={solid}] table [x=distance, y=abs_log_err, col sep=comma] {comparison_johannesburg_RYRZd1connfull.csv};
                    \addplot[black, densely dashed, mark=square, mark options={solid}] table [x=distance, y=abs_log_err, col sep=comma] {comparison_johannesburg_SwapRZd1connfull.csv};
                    
                    \nextgroupplot[title=Vigo]
                    \addplot[red, solid, mark=*, mark options={solid}] table [x=distance, y=abs_log_err, col sep=comma] {comparison_vigo_ASWAPn40m2S_z00.csv};
                    \addplot[blue, dashed, mark=o, mark options={solid}] table [x=distance, y=abs_log_err, col sep=comma] {comparison_vigo_RYd1connfull.csv};
                    \addplot[violet, dotted, mark=triangle, mark options={solid}] table [x=distance, y=abs_log_err, col sep=comma] {comparison_vigo_RYRZd1connfull.csv};
                    \addplot[black, densely dashed, mark=square, mark options={solid}] table [x=distance, y=abs_log_err, col sep=comma] {comparison_vigo_SwapRZd1connfull.csv};
                    
                    \nextgroupplot[title=Ourense]
                    \addplot[red, solid, mark=*, mark options={solid}] table [x=distance, y=abs_log_err, col sep=comma] {comparison_ourense_ASWAPn40m2S_z00.csv};
                    \addplot[blue, dashed, mark=o, mark options={solid}] table [x=distance, y=abs_log_err, col sep=comma] {comparison_ourense_RYd1connfull.csv};
                    \addplot[violet, dotted, mark=triangle, mark options={solid}] table [x=distance, y=abs_log_err, col sep=comma] {comparison_ourense_RYRZd1connfull.csv};
                    \addplot[black, densely dashed, mark=square, mark options={solid}] table [x=distance, y=abs_log_err, col sep=comma] {comparison_ourense_SwapRZd1connfull.csv};
                    
                    \end{groupplot}
                \end{tikzpicture}
                \caption{Comparison of the symmetry-preserving ASWAP ansatz with several of the best competing ad hoc ans\"atze, for a variety of noise models corresponding to real devices. In all cases, the ASWAP ansatz performs better at smaller bond distances. This is because for large bond distances, the ground state corresponds to setting all the parameters in the ad hoc ans\"atze to $0$, whereas the smaller bond distances vary from the initial state of the register more. In this sense, this is a feature only of the classical optimizer, not the ans\"atze. All results in this figure have SPAM error mitigation applied.}
                \label{fig:ans_comp}
            \end{figure*}
            
            \begin{figure*}
                \centering
                \begin{tikzpicture}
                    \begin{groupplot}[
                        group style = {group size = 2 by 4,
                        x descriptions at=edge bottom,
                        horizontal sep = 0pt,
                        vertical sep = 25pt
                        },
                        width=0.8*\columnwidth,
                        xlabel={Bond Distance (Angstroms)},
                        legend pos=north east,
                        ticklabel style={
                        /pgf/number format/fixed,
                        /pgf/number format/precision=5
                        }
                        ]
                    
                    \nextgroupplot[ylabel = $\left< N \right>$, title=Boeblingen]
                    \addplot[red, solid, mark=*, mark options={solid}] table [x=distance, y=op_mean, col sep=comma] {operator_values_N_boeblingen_ASWAPn40m2S_z00.csv};
                    \addplot[blue, dashed, mark=o, mark options={solid}] table [x=distance, y=op_mean, col sep=comma] {operator_values_N_boeblingen_RYd1connfull.csv};
                    \addplot[violet, dotted, mark=triangle, mark options={solid}] table [x=distance, y=op_mean, col sep=comma] {operator_values_N_boeblingen_RYRZd1connfull.csv};
                    \addplot[black, densely dashed, mark=square, mark options={solid}] table [x=distance, y=op_mean, col sep=comma] {operator_values_N_boeblingen_SwapRZd1connfull.csv};
                    
                    \nextgroupplot[ylabel = $\left< S_z \right>$, yticklabel pos=right, ylabel near ticks, title=Boeblingen]
                    \addplot[red, solid, mark=*, mark options={solid}] table [x=distance, y=op_mean, col sep=comma] {operator_values_S_z_boeblingen_ASWAPn40m2S_z00.csv};
                    \addplot[blue, dashed, mark=o, mark options={solid}] table [x=distance, y=op_mean, col sep=comma] {operator_values_S_z_boeblingen_RYd1connfull.csv};
                    \addplot[violet, dotted, mark=triangle, mark options={solid}] table [x=distance, y=op_mean, col sep=comma] {operator_values_S_z_boeblingen_RYRZd1connfull.csv};
                    \addplot[black, densely dashed, mark=square, mark options={solid}] table [x=distance, y=op_mean, col sep=comma] {operator_values_S_z_boeblingen_SwapRZd1connfull.csv};
                    
                    \nextgroupplot[ylabel = $\left< N \right>$, title=Johannesburg]
                    \addplot[red, solid, mark=*, mark options={solid}] table [x=distance, y=op_mean, col sep=comma] {operator_values_N_johannesburg_ASWAPn40m2S_z00.csv};
                    \addplot[blue, dashed, mark=o, mark options={solid}] table [x=distance, y=op_mean, col sep=comma] {operator_values_N_johannesburg_RYd1connfull.csv};
                    \addplot[violet, dotted, mark=triangle, mark options={solid}] table [x=distance, y=op_mean, col sep=comma] {operator_values_N_johannesburg_RYRZd1connfull.csv};
                    \addplot[black, densely dashed, mark=square, mark options={solid}] table [x=distance, y=op_mean, col sep=comma] {operator_values_N_johannesburg_SwapRZd1connfull.csv};
                    
                    \nextgroupplot[ylabel = $\left< S_z \right>$, yticklabel pos=right, ylabel near ticks, title=Johannesburg]
                    \addplot[red, solid, mark=*, mark options={solid}] table [x=distance, y=op_mean, col sep=comma] {operator_values_S_z_johannesburg_ASWAPn40m2S_z00.csv};
                    \addplot[blue, dashed, mark=o, mark options={solid}] table [x=distance, y=op_mean, col sep=comma] {operator_values_S_z_johannesburg_RYd1connfull.csv};
                    \addplot[violet, dotted, mark=triangle, mark options={solid}] table [x=distance, y=op_mean, col sep=comma] {operator_values_S_z_johannesburg_RYRZd1connfull.csv};
                    \addplot[black, densely dashed, mark=square, mark options={solid}] table [x=distance, y=op_mean, col sep=comma] {operator_values_S_z_johannesburg_SwapRZd1connfull.csv};
                    
                    \nextgroupplot[ylabel = $\left< N \right>$, title=Vigo]
                    \addplot[red, solid, mark=*, mark options={solid}] table [x=distance, y=op_mean, col sep=comma] {operator_values_N_vigo_ASWAPn40m2S_z00.csv};
                    \addplot[blue, dashed, mark=o, mark options={solid}] table [x=distance, y=op_mean, col sep=comma] {operator_values_N_vigo_RYd1connfull.csv};
                    \addplot[violet, dotted, mark=triangle, mark options={solid}] table [x=distance, y=op_mean, col sep=comma] {operator_values_N_vigo_RYRZd1connfull.csv};
                    \addplot[black, densely dashed, mark=square, mark options={solid}] table [x=distance, y=op_mean, col sep=comma] {operator_values_N_vigo_SwapRZd1connfull.csv};
                    
                    \nextgroupplot[ylabel = $\left< S_z \right>$, yticklabel pos=right, ylabel near ticks, title=Vigo]
                    \addplot[red, solid, mark=*, mark options={solid}] table [x=distance, y=op_mean, col sep=comma] {operator_values_S_z_vigo_ASWAPn40m2S_z00.csv};
                    \addplot[blue, dashed, mark=o, mark options={solid}] table [x=distance, y=op_mean, col sep=comma] {operator_values_S_z_vigo_RYd1connfull.csv};
                    \addplot[violet, dotted, mark=triangle, mark options={solid}] table [x=distance, y=op_mean, col sep=comma] {operator_values_S_z_vigo_RYRZd1connfull.csv};
                    \addplot[black, densely dashed, mark=square, mark options={solid}] table [x=distance, y=op_mean, col sep=comma] {operator_values_S_z_vigo_SwapRZd1connfull.csv};
                    
                    \nextgroupplot[ylabel = $\left< N \right>$, title=Ourense]
                    \addplot[red, solid, mark=*, mark options={solid}] table [x=distance, y=op_mean, col sep=comma] {operator_values_N_ourense_ASWAPn40m2S_z00.csv};
                    \addplot[blue, dashed, mark=o, mark options={solid}] table [x=distance, y=op_mean, col sep=comma] {operator_values_N_ourense_RYd1connfull.csv};
                    \addplot[violet, dotted, mark=triangle, mark options={solid}] table [x=distance, y=op_mean, col sep=comma] {operator_values_N_ourense_RYRZd1connfull.csv};
                    \addplot[black, densely dashed, mark=square, mark options={solid}] table [x=distance, y=op_mean, col sep=comma] {operator_values_N_ourense_SwapRZd1connfull.csv};
                    
                    \nextgroupplot[legend style={at={(1-0.03,0.5)},anchor=east}, ylabel = $\left< S_z \right>$, yticklabel pos=right, ylabel near ticks, title=Ourense]
                    \addplot[red, solid, mark=*, mark options={solid}] table [x=distance, y=op_mean, col sep=comma] {operator_values_S_z_ourense_ASWAPn40m2S_z00.csv};
                    \addplot[blue, dashed, mark=o, mark options={solid}] table [x=distance, y=op_mean, col sep=comma] {operator_values_S_z_ourense_RYd1connfull.csv};
                    \addplot[violet, dotted, mark=triangle, mark options={solid}] table [x=distance, y=op_mean, col sep=comma] {operator_values_S_z_ourense_RYRZd1connfull.csv};
                    \addplot[black, densely dashed, mark=square, mark options={solid}] table [x=distance, y=op_mean, col sep=comma] {operator_values_S_z_ourense_SwapRZd1connfull.csv};
                    
                    \legend{ASWAP, RY, RYRZ, SwapRZ}
                    
                    \end{groupplot}
                \end{tikzpicture}
                \caption{Comparisons of preserved symmetries ($N, S_z$) in ASWAP, RY, and RYRZ ans\"atze. The ASWAP ansatz consistently preserves each desired symmetry throughout the dissociation, and any deviation, in this case, is due to the noise model. On the other hand, deviations in the RY and RYRZ cases can also be attributed to the inability of the ansatz to target a particular subspace. This is most prominently shown in the case of the $S_z$ operator (bottom row). For the RY ans\"atze, the change in the resulting expectation values around $1$ Angstrom is due to the algorithm finding another, nearly degenerate solution for the ground state that has an incorrect value of $S_z$.  All results in this figure have SPAM error mitigation applied.}
                \label{fig:sym_pres}
            \end{figure*}
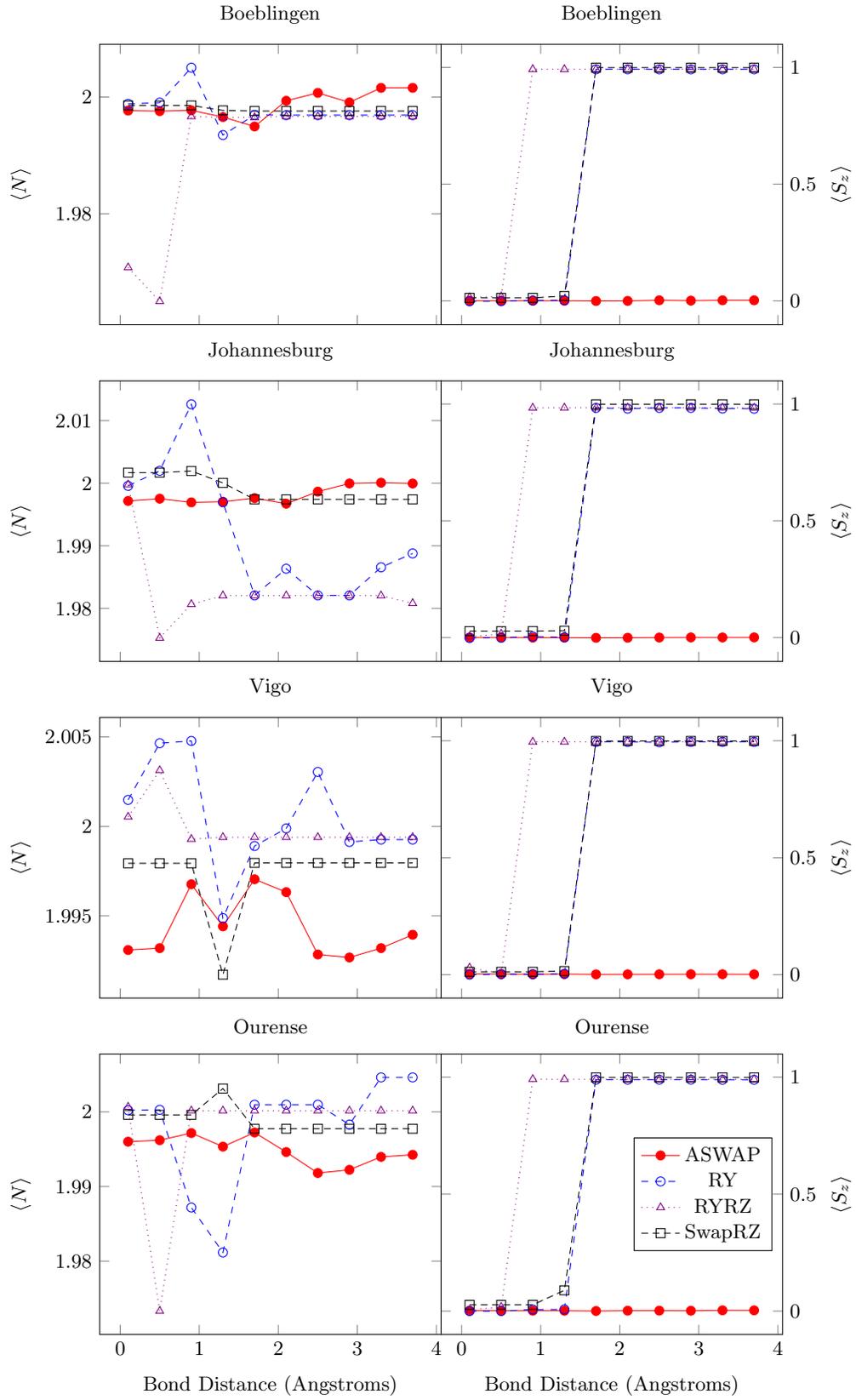
            
            In Fig.~\ref{fig:ans_comp} we show the results of running the VQE for a variety of bond distances of $\text{H}_2$ for the best-performing symmetry preserving and ad hoc ans\"atze. We find that for smaller bond distances, the ASWAP ansatz performs better in terms of finding the correct ground state energy than the ad hoc ans\"atze. At larger bond distances, it appears that the ad hoc ans\"atze perform better; however, as we show in the next section, this is actually not the case. The reason is because in this regime, the ad hoc ans\"atze are actually converging to an excited state that is nearly degenerate with the true ground state. All the ans\"atze have greater difficulty in identifying the true ground state at large bond distance because correlations are stronger here (the true ground state is a singlet). On the other hand, the excited state found by the ad hoc ans\"atze is a (triplet) product state, which is easy for the ad hoc algorithms to identify. Because ASWAP has the correct symmetry quantum numbers built in, it does not misidentify the ground state, and the error at large bond distance is larger because of the correlations in this state.

            \section{Effect of noise on symmetry preservation}\label{sec:symm_in_noisy}

            In addition to reducing computational resources and error levels, the ASWAP ansatz also correctly preserves the desired symmetry throughout the entire dissociation curve. This is illustrated in Fig.~\ref{fig:sym_pres} for the case of $\text{H}_2$. While all ans\"atze considered remain reasonably close to the correct particle number, $\left< N \right>=2$, only ASWAP succeeds in finding the right value of $\left< S_z \right>$ at all bond distances. In contrast, the RY ans\"atze find another nearly degenerate solution that has an incorrect value of $S_z$. This is a very important benefit to our ans\"atze when considering quantum chemistry problems in the sense that with the other ans\"atze, there is no way to control which solution is found when there are approximate degeneracies aside from adding penalty terms to the objective function to find different symmetry states. Furthermore, it is not clear how one could try to find a consistent solution throughout the whole curve with another ansatz. The ASWAP ansatz allows us to specify exactly which of these solutions we want to find.

            \subsection{Total Spin Ansatz Performance}
        
            So far, this work has only focused on the ASWAP ansatz from Ref.~\cite{gard2019efficient}. That work also presented another ansatz, referred to as the $E_n$ ans\"atze, where $n$ is the number of qubits. Unlike any of the other ans\"atze in the present work, it has the ability to preserve the total spin of the trial state. The $E_n$ ansatz is constructed by first writing out a general state vector in the appropriate total spin subspace, with the coefficients parameterized by hyperspherical coordinates. Then a unitary that produces these general states of fixed total spin starting from $\ket{0}^{\otimes n}$ is constructed and decomposed in terms of Toffoli gates using gray codes. The resulting ans\"atze can then be used to prepare arbitrary states with the appropriate spin projection, total spin, and particle number symmetries. The details of this construction can be found in Ref.~\cite{gard2019efficient}.

\begin{figure}[H]
                \centering
                \begin{tikzpicture}
                    \begin{axis}[
                        ylabel={Energy Error (Log Hartree)},
                        xlabel={Stretch Factor},
                        xtick={1,4,8,12,16}
                        ]
                        
                        \addplot[blue, dashed, mark=square, mark options={solid}] table [x=stretch_factor, y=log_energy_error, col sep=comma] {stretch_times_EFour.csv};
                        \addlegendentry{$E_4(S^2=0,S_z=0)$}
                        
                        \addplot[red, solid, mark=*, mark options={solid}] table [x=stretch_factor, y=log_energy_error, col sep=comma] {stretch_times_ASWAP.csv};
                        \addlegendentry{$\text{ASWAP}(S_z=0)$}
                        
                    \end{axis}
                \end{tikzpicture}
                \caption{Performance of the $E_4$ and ASWAP ans\"atze for $\text{H}_2$ at equilibrium for a range of stretch factors. The stretch factors here are the ratio of the original and simulated $T_1$ and $T_2$ times for the noise model. A stretch factor of $1$ corresponds to the original noise model. Here we chose the noise model for the IBMQ Vigo device. We find that improving the $T_1$ and $T_2$ times of the device by a factor of roughly $4$ makes the $E_4$ ansatz perform as well as the ASWAP ansatz does today. Since the only error mitigation technique used here is SPAM error mitigation, we find that roughly an order of magnitude improvement of the $T_1$ and $T_2$ times will allow the ASWAP ansatz to reach chemical accuracy without Richardson extrapolation.}
                \label{fig:e_noise}
            \end{figure}
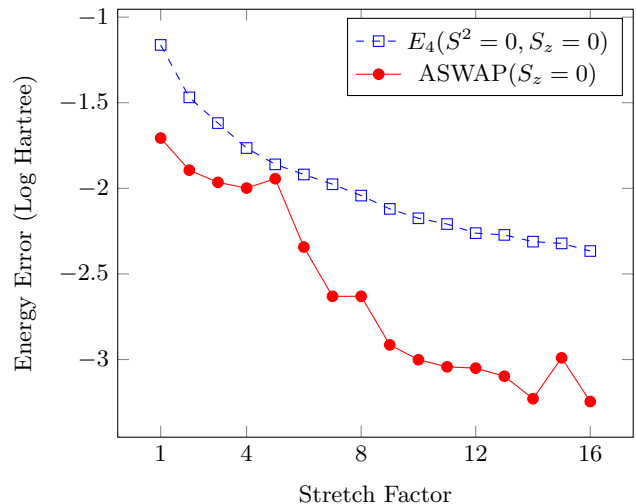

            The $E_n$ ans\"atze have been excluded from our analysis so far because they have larger gate depths, and they are generally more challenging to implement on noisy hardware. In Fig.~\ref{fig:e_noise} we compare the performance of the $E_4$ and ASWAP ans\"atze for simulating $\text{H}_2$ at equilibrium. To accomplish this, we take the noise model from the IBMQ Vigo device and uniformly stretch the $T_1$ and $T_2$ times for all of the qubits, so as to simulate a device with less noise. In doing so, we confirm that the performance of the $E_4$ gate is limited by the $T_{1/2}$ times. Moreover, we find that these times would need to be $\sim 4$ times longer than they currently are to perform as well as the ASWAP ansatz with current $T_{1/2}$ times. Hence, we expect that the $E_4$ ansatz will be useful for preserving total spin, but is not yet currently viable. Nevertheless, there may be more complex molecules for which the ability to preserve $S^2$ will have a greater impact on the results and offset the challenges of the larger circuit depth. It is also likely that there exist more efficient circuits that accomplish the same task as the $E_n$ ans\"atze. This will be explored in future work.

        \section{Excited States}\label{sec:excited}
            
            A novel aspect of our symmetry-preserving ans\"atze is the ability to find excited states by targeting subspaces of the Hilbert space that have the appropriate symmetry, provided this symmetry differs from that of the ground state. Other methods of calculating excited states using VQEs exist, but they require either iteratively running multiple VQEs to determine excited states \cite{Higgott2019variationalquantum} or executing additional circuits \cite{Colless2018, ollitrault2019quantum}. Each of these approaches has its own advantages. However, the former can lead to error accumulation as the calculation of excited states depends on the accuracy of the previously found ground state through an orthogonality condition, and the latter introduces computational overhead. In contrast, our method avoids these issues because the search for excited states does not depend on having already obtained the ground state. In fact, we could compute both ground and excited states simultaneously by running separate VQEs in parallel. Furthermore, other methods of excited-state calculations typically come with the added overhead of longer depth circuits, larger register size, and/or more required measurements. Since our method amounts to simply changing the ans\"atze for different excited states, none of these issues exist in our approach, and most excited-state calculations actually lead to shorter depth circuits. One possible disadvantage of our approach is that depending on the system chosen, it may not be the case that one can uniquely specify all excited-states by changing symmetries alone. Therefore, the optimal approach may be a combination of our techniques and some of the previously introduced ones.
            
\begin{figure*}
                \centering
                \begin{tikzpicture}
	                \begin{axis}[%
	                    minor tick num=5,
                	    legend pos=north east,
                	    enlargelimits=false,
                	    width=5in,
                	    xlabel={Bond Distance (Angstroms)},
                	    ylabel={Energy (Hartree)},
                	    xmin=0.4,
                	]
                    
                    \addplot[red, opacity=0.5, solid, mark options={solid}, forget plot] table [x=dist, y=en_plus_rep, col sep=comma] {exact_eigenenergies_0.csv};
                    \addplot[red, opacity=0.5, solid, mark options={solid}, forget plot] table [x=dist, y=en_plus_rep, col sep=comma] {exact_eigenenergies_1.csv};
                    \addplot[red, opacity=0.5, solid, mark options={solid}, forget plot] table [x=dist, y=en_plus_rep, col sep=comma] {exact_eigenenergies_2.csv};
                    \addplot[red, opacity=0.5, solid, mark options={solid}, forget plot] table [x=dist, y=en_plus_rep, col sep=comma] {exact_eigenenergies_3.csv};
                    \addplot[red, opacity=0.5, solid, mark options={solid}, forget plot] table [x=dist, y=en_plus_rep, col sep=comma] {exact_eigenenergies_4.csv};
                    \addplot[red, opacity=0.5, solid, mark options={solid}, forget plot] table [x=dist, y=en_plus_rep, col sep=comma] {exact_eigenenergies_5.csv};
                    \addplot[red, opacity=0.5, solid, mark options={solid}, forget plot] table [x=dist, y=en_plus_rep, col sep=comma] {exact_eigenenergies_6.csv};
                    \addplot[red, opacity=0.5, solid, mark options={solid}, forget plot] table [x=dist, y=en_plus_rep, col sep=comma] {exact_eigenenergies_7.csv};
                    \addplot[red, opacity=0.5, solid, mark options={solid}, forget plot] table [x=dist, y=en_plus_rep, col sep=comma] {exact_eigenenergies_8.csv};
                    \addplot[red, opacity=0.5, solid, mark options={solid}, forget plot] table [x=dist, y=en_plus_rep, col sep=comma] {exact_eigenenergies_9.csv};
                    \addplot[red, opacity=0.5, solid, mark options={solid}, forget plot] table [x=dist, y=en_plus_rep, col sep=comma] {exact_eigenenergies_10.csv};
                    \addplot[red, opacity=0.5, solid, mark options={solid}, forget plot] table [x=dist, y=en_plus_rep, col sep=comma] {exact_eigenenergies_11.csv};
                    \addplot[red, opacity=0.5, solid, mark options={solid}, forget plot] table [x=dist, y=en_plus_rep, col sep=comma] {exact_eigenenergies_12.csv};
                    \addplot[red, opacity=0.5, solid, mark options={solid}, forget plot] table [x=dist, y=en_plus_rep, col sep=comma] {exact_eigenenergies_13.csv};
                    \addplot[red, opacity=0.5, solid, mark options={solid}, forget plot] table [x=dist, y=en_plus_rep, col sep=comma] {exact_eigenenergies_14.csv};
                    \addplot[red, opacity=0.5, solid, mark options={solid}, forget plot] table [x=dist, y=en_plus_rep, col sep=comma] {exact_eigenenergies_15.csv};
                    
                    \addplot[blue, mark=square*, only marks, mark options={solid}] table [x=distance, y=vqe_plus_rep, col sep=comma] {excited_state_ASWAPn40m0S_z00.csv};
                    \addlegendentry{$\text{ASWAP}(N=0, S_z=0)$}
                    \addplot[red, only marks, mark options={solid}] table [x=distance, y=vqe_plus_rep, col sep=comma] {excited_state_ASWAPn40m1S_z05.csv};
                    \addlegendentry{$\text{ASWAP}(N=1, S_z=+1/2)$}
                    \addplot[black, mark=square*, only marks, mark options={solid}] table [x=distance, y=vqe_plus_rep, col sep=comma] {excited_state_ASWAPn40m2S_z00.csv};
                    \addlegendentry{$\text{ASWAP}(N=2, S_z=0)$}
                    \addplot[black, mark=triangle*, only marks, mark options={solid}] table [x=distance, y=vqe_plus_rep, col sep=comma] {excited_state_ASWAPn40m2S_z10.csv};
                    \addlegendentry{$\text{ASWAP}(N=2, S_z=+1)$}
                    \addplot[orange, only marks, mark options={solid}] table [x=distance, y=vqe_plus_rep, col sep=comma] {excited_state_ASWAPn40m3S_z05.csv};
                    \addlegendentry{$\text{ASWAP}(N=3, S_z=+1/2)$}
                    \addplot[purple, mark=square*, only marks, mark options={solid}] table [x=distance, y=vqe_plus_rep, col sep=comma] {excited_state_ASWAPn40m4S_z00.csv};
                    \addlegendentry{$\text{ASWAP}(N=4, S_z=0)$}
	                
                	\end{axis}
                \end{tikzpicture}
                \caption{The dissociation curve for a variety of excited states for $\text{H}_2$ calculated using the ASWAP ansatz. Dotted points correspond to results obtained using simulations of the Vigo IBMQ quantum processor, where different colors correspond to different symmetries enforced by the ASWAP ansatz. Lines correspond to the different excited states as a function of the bond distance calculated using exact diagonalization, where different colors correspond to different excited states. In all cases, we find strong agreement between the results from noisy simulation and exact diagonalization.  All results in this figure have SPAM error mitigation applied.}
                \label{fig:excited}
            \end{figure*}
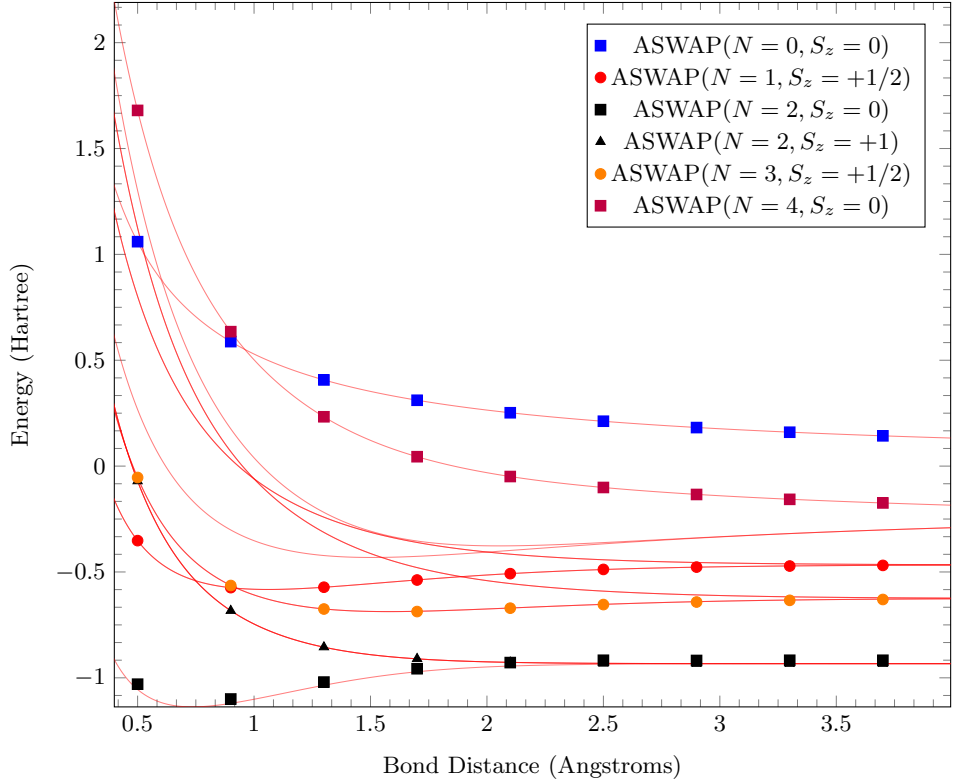

            A demonstration of excited state calculations using our ASWAP ansatz is shown for the $\text{H}_2$ molecule for several excited states in Fig.~\ref{fig:excited}. Note that some of the exact results do not have corresponding simulation results. This is because our method does not, in general, allow one to produce all of the higher excited states, only those directly accessible by partitioning the Hilbert space according to symmetries. In all cases considered, we are able to prepare $5$ excited states (in addition to the ground state) out of the possible $10$ (not counting degeneracies). The error rates here are sufficiently low so that one may correctly order the eigenstates according to their energies.
            
        \section{Conclusions}\label{sec:conc}
        
            In this work, we have demonstrated that the deleterious effects of noise on a VQE can be mitigated by exploiting the symmetries of the Hamiltonian being simulated. We achieve this by performing simulations of the noiseless and noisy systems for a variety of ans\"atze, utilizing the most current error mitigation techniques. The results of these simulations indicate that, with or without noise, using state preparation circuits that preserve the relevant symmetries of the problem reduces computational resources and error rates compared to more ad hoc approaches. In addition, we have shown that the built-in symmetry preservation of our ans\"atze allows us to find excited states without having to first find the ground state and without using longer state preparation circuits.
            
            \section*{ACKNOWLEDGEMENTS}
S. E. E. acknowledges support from the US Department of Energy (Award No. DE-SC0019318). E. B. and N. J. M. acknowledge support from the US Department of Energy (Award No. DE-SC0019199). This research used quantum computing system resources supported by the U.S. Department of Energy, Office of Science, Office of Advanced Scientific Computing Research program office. Oak Ridge National Laboratory manages access to the IBM Q System as part of the IBM Q Network.
            
\clearpage

\bibliographystyle{unsrt}
\bibliography{bibliography}

\begin{thebibliography}{10}

\bibitem{Feynman1982}
Richard~P. Feynman.
\newblock Simulating physics with computers.
\newblock {\em International Journal of Theoretical Physics}, 21(6):467--488,
  Jun 1982.

\bibitem{Peruzzo2014}
A.~Peruzzo, J.McClean, P.~Shadbolt, M.-H.Yung, X.-Q. Zhou, P.J. Love,
  A.~Aspuru-Guzik, and J.~L. O’Brien.
\newblock A variational eigenvalue solver on a photonic quantum processor.
\newblock {\em Nature Commun.}, 5(4213):4213, 2014.

\bibitem{OMalley2016}
P.~J.~J. O'Malley, R.~Babbush, I.~D. Kivlichan, J.~Romero, J.~R. McClean,
  R.~Barends, J.~Kelly, P.~Roushan, A.~Tranter, N.~Ding, B.~Campbell, Y.~Chen,
  Z.~Chen, B.~Chiaro, A.~Dunsworth, A.~G. Fowler, E.~Jeffrey, E.~Lucero,
  A.~Megrant, J.~Y. Mutus, M.~Neeley, C.~Neill, C.~Quintana, D.~Sank,
  A.~Vainsencher, J.~Wenner, T.~C. White, P.~V. Coveney, P.~J. Love, H.~Neven,
  A.~Aspuru-Guzik, and J.~M. Martinis.
\newblock Scalable quantum simulation of molecular energies.
\newblock {\em Phys. Rev. X}, 6:031007, Jul 2016.

\bibitem{McClean2016}
Jarrod~R McClean, Jonathan Romero, Ryan Babbush, and Al{\'{a}}n Aspuru-Guzik.
\newblock The theory of variational hybrid quantum-classical algorithms.
\newblock {\em New J. Phys}, 18(2):023023, feb 2016.

\bibitem{kandala2017hardware}
Abhinav Kandala, Antonio Mezzacapo, Kristan Temme, Maika Takita, Markus Brink,
  Jerry~M Chow, and Jay~M Gambetta.
\newblock Hardware-efficient variational quantum eigensolver for small
  molecules and quantum magnets.
\newblock {\em Nature}, 549(7671):242--246, 2017.

\bibitem{Colless2018}
J.~I. Colless, V.~V. Ramasesh, D.~Dahlen, M.~S. Blok, M.~E. Kimchi-Schwartz,
  J.~R. McClean, J.~Carter, W.~A. de~Jong, and I.~Siddiqi.
\newblock Computation of molecular spectra on a quantum processor with an
  error-resilient algorithm.
\newblock {\em Phys. Rev. X}, 8:011021, Feb 2018.

\bibitem{Preskill2018}
John Preskill.
\newblock Quantum {C}omputing in the {NISQ} era and beyond.
\newblock {\em {Quantum}}, 2:79, August 2018.

\bibitem{farhi2014quantum}
Edward Farhi, Jeffrey Goldstone, and Sam Gutmann.
\newblock A quantum approximate optimization algorithm.
\newblock {\em arXiv preprint arXiv:1411.4028}, 2014.

\bibitem{pagano2019quantum}
G~Pagano, A~Bapat, P~Becker, KS~Collins, A~De, PW~Hess, HB~Kaplan,
  A~Kyprianidis, WL~Tan, C~Baldwin, et~al.
\newblock Quantum approximate optimization with a trapped-ion quantum
  simulator.
\newblock {\em arXiv preprint arXiv:1906.02700}, 2019.

\bibitem{romero2019variational}
Jonathan Romero and Alan Aspuru-Guzik.
\newblock Variational quantum generators: Generative adversarial quantum
  machine learning for continuous distributions.
\newblock {\em arXiv preprint arXiv:1901.00848}, 2019.

\bibitem{leyton2019robust}
Vicente Leyton-Ortega, Alejandro Perdomo-Ortiz, and Oscar Perdomo.
\newblock Robust implementation of generative modeling with parametrized
  quantum circuits.
\newblock {\em arXiv preprint arXiv:1901.08047}, 2019.

\bibitem{Zhueaaw9918}
D.~Zhu, N.~M. Linke, M.~Benedetti, K.~A. Landsman, N.~H. Nguyen, C.~H.
  Alderete, A.~Perdomo-Ortiz, N.~Korda, A.~Garfoot, C.~Brecque, L.~Egan,
  O.~Perdomo, and C.~Monroe.
\newblock Training of quantum circuits on a hybrid quantum computer.
\newblock {\em Science Advances}, 5(10), 2019.

\bibitem{benedetti2019generative}
Marcello Benedetti, Delfina Garcia-Pintos, Oscar Perdomo, Vicente
  Leyton-Ortega, Yunseong Nam, and Alejandro Perdomo-Ortiz.
\newblock A generative modeling approach for benchmarking and training shallow
  quantum circuits.
\newblock {\em npj Quantum Information}, 5(1):1--9, 2019.

\bibitem{verdon2019learning}
Guillaume Verdon, Michael Broughton, Jarrod~R McClean, Kevin~J Sung, Ryan
  Babbush, Zhang Jiang, Hartmut Neven, and Masoud Mohseni.
\newblock Learning to learn with quantum neural networks via classical neural
  networks.
\newblock {\em arXiv preprint arXiv:1907.05415}, 2019.

\bibitem{Kivlichan2018}
Ian~D. Kivlichan, Jarrod McClean, Nathan Wiebe, Craig Gidney, Al\'an
  Aspuru-Guzik, Garnet Kin-Lic Chan, and Ryan Babbush.
\newblock Quantum simulation of electronic structure with linear depth and
  connectivity.
\newblock {\em Phys. Rev. Lett.}, 120:110501, Mar 2018.

\bibitem{PhysRevA.99.062304}
Tyson Jones, Suguru Endo, Sam McArdle, Xiao Yuan, and Simon~C. Benjamin.
\newblock Variational quantum algorithms for discovering hamiltonian spectra.
\newblock {\em Phys. Rev. A}, 99:062304, Jun 2019.

\bibitem{PhysRevApplied.11.044087}
Kosuke Mitarai, Tennin Yan, and Keisuke Fujii.
\newblock Generalization of the output of a variational quantum eigensolver by
  parameter interpolation with a low-depth ansatz.
\newblock {\em Phys. Rev. Applied}, 11:044087, Apr 2019.

\bibitem{PhysRevLett.122.230401}
Robert~M. Parrish, Edward~G. Hohenstein, Peter~L. McMahon, and Todd~J.
  Mart\'{\i}nez.
\newblock Quantum computation of electronic transitions using a variational
  quantum eigensolver.
\newblock {\em Phys. Rev. Lett.}, 122:230401, Jun 2019.

\bibitem{anschuetz2019variational}
Eric Anschuetz, Jonathan Olson, Al{\'a}n Aspuru-Guzik, and Yudong Cao.
\newblock Variational quantum factoring.
\newblock In {\em International Workshop on Quantum Technology and Optimization
  Problems}, pages 74--85. Springer, 2019.

\bibitem{Moll_2018}
Nikolaj Moll, Panagiotis Barkoutsos, Lev~S Bishop, Jerry~M Chow, Andrew Cross,
  Daniel~J Egger, Stefan Filipp, Andreas Fuhrer, Jay~M Gambetta, Marc Ganzhorn,
  Abhinav Kandala, Antonio Mezzacapo, Peter Müller, Walter Riess, Gian Salis,
  John Smolin, Ivano Tavernelli, and Kristan Temme.
\newblock Quantum optimization using variational algorithms on near-term
  quantum devices.
\newblock {\em Quantum Science and Technology}, 3(3):030503, jun 2018.

\bibitem{Barkoutsos2018}
Panagiotis~Kl. Barkoutsos, Jerome~F. Gonthier, Igor Sokolov, Nikolaj Moll, Gian
  Salis, Andreas Fuhrer, Marc Ganzhorn, Daniel~J. Egger, Matthias Troyer,
  Antonio Mezzacapo, Stefan Filipp, and Ivano Tavernelli.
\newblock Quantum algorithms for electronic structure calculations:
  Particle-hole hamiltonian and optimized wave-function expansions.
\newblock {\em Phys. Rev. A}, 98:022322, Aug 2018.

\bibitem{gard2019efficient}
Bryan~T Gard, Linghua Zhu, George~S Barron, Nicholas~J Mayhall, Sophia~E
  Economou, and Edwin Barnes.
\newblock Efficient symmetry-preserving state preparation circuits for the
  variational quantum eigensolver algorithm.
\newblock {\em arXiv preprint arXiv:1904.10910}, 2019.

\bibitem{doi:10.1002/qute.201900070}
Sukin Sim, Peter~D. Johnson, and Alán Aspuru-Guzik.
\newblock Expressibility and entangling capability of parameterized quantum
  circuits for hybrid quantum-classical algorithms.
\newblock {\em Advanced Quantum Technologies}, 2(12):1900070, 2019.

\bibitem{barkoutsos2019improving}
Panagiotis~Kl Barkoutsos, Giacomo Nannicini, Anton Robert, Ivano Tavernelli,
  and Stefan Woerner.
\newblock Improving variational quantum optimization using cvar.
\newblock {\em arXiv preprint arXiv:1907.04769}, 2019.

\bibitem{robert2019resource}
Anton Robert, Panagiotis~Kl Barkoutsos, Stefan Woerner, and Ivano Tavernelli.
\newblock Resource-efficient quantum algorithm for protein folding.
\newblock {\em arXiv preprint arXiv:1908.02163}, 2019.

\bibitem{PhysRevA.99.012334}
Suguru Endo, Qi~Zhao, Ying Li, Simon Benjamin, and Xiao Yuan.
\newblock Mitigating algorithmic errors in a hamiltonian simulation.
\newblock {\em Phys. Rev. A}, 99:012334, Jan 2019.

\bibitem{larose2019variational}
Ryan LaRose, Arkin Tikku, {\'E}tude O’Neel-Judy, Lukasz Cincio, and Patrick~J
  Coles.
\newblock Variational quantum state diagonalization.
\newblock {\em npj Quantum Information}, 5(1):1--10, 2019.

\bibitem{PhysRevLett.120.210501}
E.~F. Dumitrescu, A.~J. McCaskey, G.~Hagen, G.~R. Jansen, T.~D. Morris,
  T.~Papenbrock, R.~C. Pooser, D.~J. Dean, and P.~Lougovski.
\newblock Cloud quantum computing of an atomic nucleus.
\newblock {\em Phys. Rev. Lett.}, 120:210501, May 2018.

\bibitem{PhysRevA.98.032331}
N.~Klco, E.~F. Dumitrescu, A.~J. McCaskey, T.~D. Morris, R.~C. Pooser, M.~Sanz,
  E.~Solano, P.~Lougovski, and M.~J. Savage.
\newblock Quantum-classical computation of schwinger model dynamics using
  quantum computers.
\newblock {\em Phys. Rev. A}, 98:032331, Sep 2018.

\bibitem{kokail2019self}
Christian Kokail, Christine Maier, Rick van Bijnen, Tiff Brydges, Manoj~K
  Joshi, Petar Jurcevic, Christine~A Muschik, Pietro Silvi, Rainer Blatt,
  Christian~F Roos, et~al.
\newblock Self-verifying variational quantum simulation of lattice models.
\newblock {\em Nature}, 569(7756):355--360, 2019.

\bibitem{Higgott2019variationalquantum}
Oscar Higgott, Daochen Wang, and Stephen Brierley.
\newblock Variational {Q}uantum {C}omputation of {E}xcited {S}tates.
\newblock {\em {Quantum}}, 3:156, July 2019.

\bibitem{ollitrault2019quantum}
Pauline~J Ollitrault, Abhinav Kandala, Chun-Fu Chen, Panagiotis~Kl Barkoutsos,
  Antonio Mezzacapo, Marco Pistoia, Sarah Sheldon, Stefan Woerner, Jay
  Gambetta, and Ivano Tavernelli.
\newblock Quantum equation of motion for computing molecular excitation
  energies on a noisy quantum processor.
\newblock 2019.

\bibitem{Grimsley2018}
Harper~R. Grimsley, Sophia~E. Economou, Edwin Barnes, and Nicholas~J. Mayhall.
\newblock An adaptive variational algorithm for exact molecular simulations on
  a quantum computer.
\newblock {\em Nat. Commun.}, 10:3007, 2019.

\bibitem{tang2019qubitadaptvqe}
Ho~Lun Tang, Edwin Barnes, Harper~R. Grimsley, Nicholas~J. Mayhall, and
  Sophia~E. Economou.
\newblock qubit-adapt-vqe: An adaptive algorithm for constructing
  hardware-efficient ansatze on a quantum processor, 2019.

\bibitem{Kuhn2018}
Michael Kuehn, Sebastian Zanker, Peter Deglmann, Michael Marthaler, and Horst
  Weiss.
\newblock Accuracy and resource estimations for quantum chemistry on a
  near-term quantum computer.
\newblock {\em Journal of Chemical Theory and Computation}, 0(ja):null, 0.
\newblock PMID: 31403781.

\bibitem{McClean2018}
Jarrod~R. McClean, Sergio Boixo, Vadim~N. Smelyanskiy, Ryan Babbush, and
  Hartmut Neven.
\newblock Barren plateaus in quantum neural network training landscapes.
\newblock {\em Nat. Commun.}, 9:4812, 2018.

\bibitem{Grant2019initialization}
Edward Grant, Leonard Wossnig, Mateusz Ostaszewski, and Marcello Benedetti.
\newblock An initialization strategy for addressing barren plateaus in
  parametrized quantum circuits.
\newblock {\em {Quantum}}, 3:214, December 2019.

\bibitem{parrish2019jacobi}
Robert~M Parrish, Joseph~T Iosue, Asier Ozaeta, and Peter~L McMahon.
\newblock A jacobi diagonalization and anderson acceleration algorithm for
  variational quantum algorithm parameter optimization.
\newblock {\em arXiv preprint arXiv:1904.03206}, 2019.

\bibitem{parrish2019hybrid}
Robert~M Parrish, Edward~G Hohenstein, Peter~L McMahon, and Todd~J Martinez.
\newblock Hybrid quantum/classical derivative theory: Analytical gradients and
  excited-state dynamics for the multistate contracted variational quantum
  eigensolver.
\newblock {\em arXiv preprint arXiv:1906.08728}, 2019.

\bibitem{verteletskyi2019measurement}
Vladyslav Verteletskyi, Tzu-Ching Yen, and Artur~F Izmaylov.
\newblock Measurement optimization in the variational quantum eigensolver using
  a minimum clique cover.
\newblock {\em arXiv preprint arXiv:1907.03358}, 2019.

\bibitem{doi:10.1021/acs.jctc.9b00791}
Artur~F. Izmaylov, Tzu-Ching Yen, Robert~A. Lang, and Vladyslav Verteletskyi.
\newblock Unitary partitioning approach to the measurement problem in the
  variational quantum eigensolver method.
\newblock {\em Journal of Chemical Theory and Computation}, 16(1):190--195,
  2020.
\newblock PMID: 31747266.

\bibitem{gokhale2019minimizing}
Pranav Gokhale, Olivia Angiuli, Yongshan Ding, Kaiwen Gui, Teague Tomesh,
  Martin Suchara, Margaret Martonosi, and Frederic~T Chong.
\newblock Minimizing state preparations in variational quantum eigensolver by
  partitioning into commuting families.
\newblock {\em arXiv preprint arXiv:1907.13623}, 2019.

\bibitem{wilson2019optimizing}
Max Wilson, Sam Stromswold, Filip Wudarski, Stuart Hadfield, Norm~M Tubman, and
  Eleanor Rieffel.
\newblock Optimizing quantum heuristics with meta-learning.
\newblock {\em arXiv preprint arXiv:1908.03185}, 2019.

\bibitem{bonet2019nearly}
Xavier Bonet-Monroig, Ryan Babbush, and Thomas~E O'Brien.
\newblock Nearly optimal measurement scheduling for partial tomography of
  quantum states.
\newblock {\em arXiv preprint arXiv:1908.05628}, 2019.

\bibitem{izmaylov2019revising}
Artur~F Izmaylov, Tzu-Ching Yen, and Ilya~G Ryabinkin.
\newblock Revising the measurement process in the variational quantum
  eigensolver: is it possible to reduce the number of separately measured
  operators?
\newblock {\em Chemical science}, 10(13):3746--3755, 2019.

\bibitem{Bonet2018}
X.~Bonet-Monroig, R.~Sagastizabal, M.~Singh, and T.~E. O'Brien.
\newblock Low-cost error mitigation by symmetry verification.
\newblock {\em Phys. Rev. A}, 98:062339, Dec 2018.

\bibitem{McArdle2018}
Sam McArdle, Xiao Yuan, and Simon Benjamin.
\newblock Error-mitigated digital quantum simulation.
\newblock {\em Phys. Rev. Lett.}, 122:180501, May 2019.

\bibitem{PhysRevA.100.010302}
R.~Sagastizabal, X.~Bonet-Monroig, M.~Singh, M.~A. Rol, C.~C. Bultink, X.~Fu,
  C.~H. Price, V.~P. Ostroukh, N.~Muthusubramanian, A.~Bruno, M.~Beekman,
  N.~Haider, T.~E. O'Brien, and L.~DiCarlo.
\newblock Experimental error mitigation via symmetry verification in a
  variational quantum eigensolver.
\newblock {\em Phys. Rev. A}, 100:010302, Jul 2019.

\bibitem{PhysRevX.10.011004}
Tyler Takeshita, Nicholas~C. Rubin, Zhang Jiang, Eunseok Lee, Ryan Babbush, and
  Jarrod~R. McClean.
\newblock Increasing the representation accuracy of quantum simulations of
  chemistry without extra quantum resources.
\newblock {\em Phys. Rev. X}, 10:011004, Jan 2020.

\bibitem{qiskit}
https://github.com/qiskit/qiskit.
\newblock Qiskit: An open-source framework for quantum computing, 2019.

\bibitem{Jordan1928}
P.~Jordan and E.~Wigner.
\newblock \"uber das paulische \"aquivalenzverbot.
\newblock {\em Z. Phys.}, 47:631, 1928.

\bibitem{doi:10.1063/1.4768229}
Jacob~T. Seeley, Martin~J. Richard, and Peter~J. Love.
\newblock The bravyi-kitaev transformation for quantum computation of
  electronic structure.
\newblock {\em The Journal of Chemical Physics}, 137(22):224109, 2012.

\bibitem{BRAVYI2002210}
Sergey~B. Bravyi and Alexei~Yu. Kitaev.
\newblock Fermionic quantum computation.
\newblock {\em Annals of Physics}, 298(1):210 -- 226, 2002.

\bibitem{PhysRevResearch.1.033033}
Kanav Setia, Sergey Bravyi, Antonio Mezzacapo, and James~D. Whitfield.
\newblock Superfast encodings for fermionic quantum simulation.
\newblock {\em Phys. Rev. Research}, 1:033033, Oct 2019.

\bibitem{PhysRevApplied.11.044092}
M.~Ganzhorn, D.J. Egger, P.~Barkoutsos, P.~Ollitrault, G.~Salis, N.~Moll,
  M.~Roth, A.~Fuhrer, P.~Mueller, S.~Woerner, I.~Tavernelli, and S.~Filipp.
\newblock Gate-efficient simulation of molecular eigenstates on a quantum
  computer.
\newblock {\em Phys. Rev. Applied}, 11:044092, Apr 2019.

\bibitem{PhysRevLett.114.200501}
Easwar Magesan, Jay~M. Gambetta, A.~D. C\'orcoles, and Jerry~M. Chow.
\newblock Machine learning for discriminating quantum measurement trajectories
  and improving readout.
\newblock {\em Phys. Rev. Lett.}, 114:200501, May 2015.

\bibitem{sun2018efficient}
Mingyu Sun and Michael~R. Geller.
\newblock Efficient characterization of correlated spam errors.
\newblock 2018.

\bibitem{PhysRevLett.119.180509}
Kristan Temme, Sergey Bravyi, and Jay~M. Gambetta.
\newblock Error mitigation for short-depth quantum circuits.
\newblock {\em Phys. Rev. Lett.}, 119:180509, Nov 2017.

\bibitem{kandala2019error}
Abhinav Kandala, Kristan Temme, Antonio~D C{\'o}rcoles, Antonio Mezzacapo,
  Jerry~M Chow, and Jay~M Gambetta.
\newblock Error mitigation extends the computational reach of a noisy quantum
  processor.
\newblock {\em Nature}, 567(7749):491--495, 2019.

\bibitem{spall1992multivariate}
James~C. Spall.
\newblock Multivariate stochastic approximation using a simultaneous
  perturbation gradient approximation.
\newblock {\em IEEE transactions on automatic control}, 37(3):332--341, 1992.

\bibitem{kingma2014adam}
Diederik~P. Kingma and Jimmy Ba.
\newblock Adam: A method for stochastic optimization.
\newblock 2014.

\bibitem{Powell1994}
M.~J.~D. Powell.
\newblock {\em A Direct Search Optimization Method That Models the Objective
  and Constraint Functions by Linear Interpolation}.
\newblock Springer Netherlands, Dordrecht, 1994.

\bibitem{Jones1993}
D.~R. Jones, C.~D. Perttunen, and B.~E. Stuckman.
\newblock Lipschitzian optimization without the lipschitz constant.
\newblock {\em Journal of Optimization Theory and Applications},
  79(1):157--181, Oct 1993.

\end{thebibliography}

\end{document}